\documentstyle [psfig]{mn}

\begin{document}
\setcounter{secnumdepth}{2}

\title[The Galaxy Environment of a Quasar at $z=1.226$]
{The Galaxy Environment of a Quasar at z=1.226: \\ A Possible Cluster Merger}

\author[C.P. Haines, R.G. Clowes, L.E. Campusano and A.J. Adamson]
{C.P. Haines$^{1,2,}$\thanks{Visiting astronomers, Cerro Tololo Inter-American Observatory, National Optical Astronomy Observatories, which are operated by the Association of Universities for Research in Astronomy, Inc. under contract with the National Science Foundation.},
 R.G. Clowes$^{1,}$\footnotemark[1], L.E. Campusano$^{2}$ and A.J. Adamson$^{3}$\\
1. Centre for Astrophysics, University of Central Lancashire, Preston PR1 2HE\\
2. Observatorio Astron\'{o}mico Cerro Cal\'{a}n, Departamento de 
Astronom\'{i}a, Universidad de Chile, Casilla 36-D, Santiago, Chile. \\
3. UKIRT, Joint Astronomy Centre, 660 North A'ohoku Place, University Park, Hilo, HI 96720, USA}

\maketitle
\def\mpc {h^{-1} {\rm{Mpc}}}
\def\kpc {h^{-1} {\rm{kpc}}}
\def\etal {et al. }
\def\rmd {\rm d}
\def\gsim{ \lower .75ex \hbox{$\sim$} \llap{\raise .27ex \hbox{$>$}}}
\def\lsim{ \lower .75ex \hbox{$\sim$} \llap{\raise .27ex \hbox{$<$}}}
\def\xvec{ \mbox{\boldmath $x$}}

\begin{abstract}

We have conducted ultra-deep optical and deep near-infrared observations of a field around the $z=1.226$ radio-quiet quasar 104420.8+055739 from the Clowes-Campusano LQG of 18 quasars at $z\sim1.3$ in search of associated galaxy clustering. Galaxies at these redshifts are distinguished by their extremely-red colours, with $I-K>3.75$, and we find a factor $\sim$11 overdensity of such galaxies in a $2.25\times2.25\,{\rm arcmin}^{2}$ field centred on the quasar. In particular, we find 15--18 galaxies with colours consistent with being a population of passively-evolving massive ellipticals at the quasar redshift. They form `fingers' in the $V-K/K$, $I-K/K$ colour-magnitude plots at $V-K\simeq6.9$, $I-K\simeq4.3$ comparable to the red sequences observed in other $z\simeq1.2$ clusters. We find suggestive evidence for substructure among the red sequence galaxies in the $K$ image, in the form of two compact groups, 40 arcsec to the north, and 60 arcsec to the south-east of the quasar. An examination of the wider optical images indicates that this substructure is significant, and that the clustering extends to form a large-scale structure  2--3 $\mpc$ across.
 We find evidence for a high ($\gsim50\%$) fraction of blue galaxies in this system, in the form of 15--20 `red-outlier' galaxies with $I-K>3.75$ and $V-I<2.00$, which we suggest are dusty, star-forming galaxies at the quasar redshift.
 Within 30 arcsec of the quasar we find a concentration of blue ($V-I<1$) galaxies in a band that bisects the two groups of red sequence galaxies. This band of blue galaxies is presumed to correspond to a region of enhanced star-formation. We explain this distribution of galaxies as the early-stages of a cluster merger which has triggered both the star-formation and the quasar.  
\end{abstract}


\section{Introduction}

Quasars are sparsely distributed across the universe, with the majority 
appearing to be unrelated to one another. However quasar clustering has been observed on large scales in the form of Large Quasar Groups (LQGs) (e.g. Webster 1982; Crampton, Cowley \& Hartwick 1987,1989; Clowes \& Campusano 1991,1994; Graham, Clowes \& Campusano 1995), where 4--25 quasars form structures 100--200$\mpc$ across. They are thus comparable in size to the largest structures seen at the present epoch, such as the `Great Wall' of galaxies, but are seen at an earlier epoch with \mbox{$0.4\lsim z\lsim2$}. It has been suggested \cite{komberg94} that LQGs represent the progenitors of these local large-scale structures, not only because of their comparable sizes, but also because of a claim that the comoving number densities of LQGs and local superclusters are comparable \cite{komberg96}.

Observations of the galaxy environments of individual quasars lends credence to the hypothesis that LQGs trace superclusters. Quasars at $z\sim0.5$ are known to favour young, low to moderately rich clusters, with richness increasing with both redshift and radio-loudness \cite{ellingson91}. At higher redshifts, a survey of 31 \mbox{$1\lsim z\lsim 2$} radio-loud quasars \cite{hall98} indicates that they are on average found in moderately rich ($R_{Abell}\approx1.5$) clusters.
 Optical and narrow-band (O\,{\sc ii}) observations of fields centred on 11 $z\sim1.1$ quasars from the Crampton \etal \shortcite{crampton89} LQG show an excess of galaxies around all but one of the quasars (Hutchings, Crampton \& Johnson 1995, hereafter HCJ; Hutchings, Crampton \& Persram 1993, hereafter HCP). In particular they find excesses of blue and emission-line galaxies around several of the quasars indicating regions of significant star-formation, a phenomenon probably related to the fact that both quasars and star-formation require the disruption of the galaxy causing a collapse of significant amounts of gas. They find no evidence for a coherent structure connecting the quasars, but star-formation is likely to be localised, rather than occurring simultaneously across large scales, as a cluster mass halo collapses or during a cluster merger.

To determine whether LQGs trace superclusters requires observations capable of identifying the quiescent galaxies that mark out any associated supercluster, in particular the massive ellipticals which dominate cluster cores. These are both the most luminous and the reddest galaxies in nearby clusters, and form a homogeneous population with very tight linear colour-magnitude (C-M) relations known as {\em red sequences} \cite{bower92}, indicative of old (12--13 Gyr) stellar populations.
 This red sequence has been followed in the optical for clusters out to $z\sim0.9$ (Arag\'{o}n-Salamanca \etal 1993; Stanford, Eisenhardt \& Dickinson 1998) and is consistent with the passive evolution of galaxies that formed in a monolithic collapse at $z\gsim3$ \cite{eggen62}. At $z\gsim1$ these galaxies should be characterised by extremely-red optical-NIR colours ($I-K\simeq4$) as the 4000\AA$\,$ break is redshifted into the $I$ band. Good contrasts of $z\gsim1$ clusters over the background are possible as field $I-K\simeq4$ galaxies appear rare: the Hawaii K-band survey finds no galaxies with $I-K>4$ for $K<18$ over an area of $86.7\,{\rm arcmin}^{2}$ (Cowie \etal 1994; Songaila \etal 1994).

Several $z\simeq1.2$ clusters have been found by searching for galaxies with $I-K\simeq4$ in fields around both targeted high-redshift AGN (Dickinson 1995, hereafter D95; Yamada \etal 1997; Tanaka \etal 2000a, hereafter T00a) and regions of extended X-ray emission (Stanford \etal 1997, hereafter S97; Rosati \etal 1999, hereafter R99). These galaxies are found to have the optical-NIR colours ($I-K\simeq4, R-K\simeq6$) expected for passively-evolving galaxies which are 2--3 Gyr old, in good agreement with the monolithic collapse model predictions. D95 also finds a red sequence for the $z=1.2$ cluster around the radio-galaxy 3C324 with $R-K\simeq5.9$ and an rms scatter of only 0.07 mag, suggesting that these galaxies formed within 300 Myr of one another. 

\begin{figure}
\psfig{figure=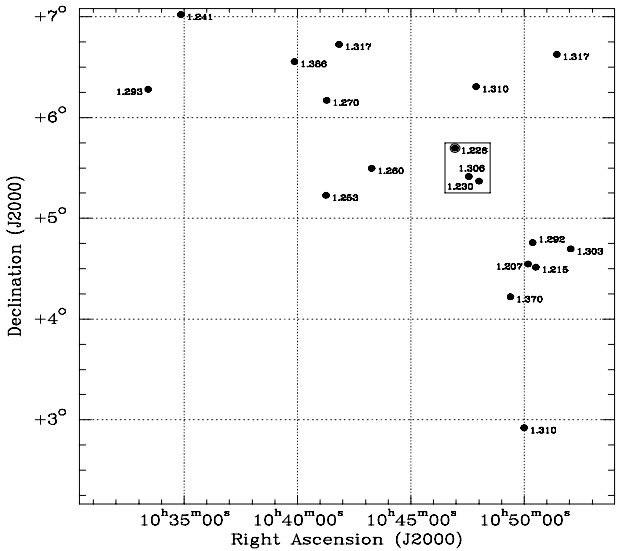,width=4.4in}
\caption{The Clowes-Campusano LQG with each quasar labelled by its redshift. The \mbox{$30\times30\,{\rm arcmin}^{2}$} region targeted for BTC imaging is indicated by the box, and the quasar for which we have $K$ imaging is circled. The boundaries of the plot match the boundaries of the AQD survey of ESO/SERC field 927 (Clowes \& Campusano 1991,1994; Clowes, Campusano \& Graham 1999). Note that this plot and Figures~\ref{spatial}, \ref{densitymap} and \ref{lss} have east to the right.}
\label{LQG}
\end{figure}

In this study we aim to establish the galaxy environment of LQGs in a manner unbiased with respect to galaxy type. We have targeted a \mbox{$30\times30\,{\rm arcmin}^{2}$} field containing three quasars from the $z\simeq1.3$ Clowes \& Campusano LQG (Fig.~\ref{LQG}) for ultra-deep optical observations complemented with deep NIR observations of selected subfields, capable of detecting the passively-evolving galaxies that should mark out any coherent large-scale structures. We present here the first results from this study, using ultra-deep $V$ and $I$ images and a $K$ image centred on the z=1.226 radio-quiet (not detected at the 1mJy level by the FIRST VLA 20cm survey) quasar 104420.8+055739. We have detected clustering of galaxies with the extremely-red colours expected of quiescent ellipticals at the quasar redshift. Throughout the paper we adopt $q_{0}=0.5$ and \mbox{${\rm H}_{0}=100h\,{\rm km\,s}^{-1}\,{\rm Mpc}^{-1}$}.

\section{Observations}

The \mbox{$30\times30\,{\rm arcmin}^{2}$} field was observed through $V$ and $I$ filters using the Big Throughput Camera (BTC) on the \mbox{4-m} Blanco telescope at the Cerro Tololo Inter-American Observatory on April 21/22 and 22/23 1998. The BTC is made up of 4 $2048\times2048$ CCDs which have pixels of size 0.43 arcsec giving a field of view for each CCD of \mbox{$14.7\times14.7\,{\rm arcmin}^{2}$}. The CCDs are arranged in a 2 by 2 grid and are separated by gaps of 5.6 arcmin. To obtain a contiguous image it was necessary to shift the telescope between exposures. 16 exposures arranged in a 4 by 4 grid, with each pointing offset by the gap width to adjacent pointings, were found to give a contiguous and almost uniform coverage over a \mbox{$30\times30\,{\rm arcmin}^{2}$} field, with $\sim90\%$ of the field covered by exactly 9 of the 16 exposures. The images had large relative offsets, and so that they could be registered, the effect of the distortions (produced by the camera optics) were removed using models based on observations of astrometric fields. These distortions are significant due to the wide field of the camera, and were of the order of 60 pixels at the corners of the CCD array. After the removal of distortions, the rms errors of registration between images were typically $\sim 0.05$ arcsec. Details of the observing and reduction procedures for the whole BTC image will be described elsewhere. Most of the reduction processes were performed in the usual way using {\sc iraf} tools.

$K$ imaging was obtained for a \mbox{$2.25\times2.25\,{\rm arcmin}^{2}$} field centred on the radio-quiet quasar at $10^{h}46^{m}56.70^{s}$, $+05^{\circ}41'50.5''$(J2000) using the UFTI camera on the \mbox{3.8-m} UKIRT telescope in March 1999. This has a field of \mbox{$1.5\times1.5\,{\rm arcmin}^{2}$} with a pixel size of 0.09 arcsec. The field around this quasar was observed in preference to the other two because of its more populous immediate environment (within 5 arcsec), suggesting that it was the most likely to be in an interacting system. No cluster-scale environmental factors were considered, and so, for the overall clustering statistics, the selection can be reasonably regarded as random.
 A standard 9 point jitter pattern was used, with each 1 minute exposure offset by 20 arcsec relative to adjacent exposures. A dark frame was obtained between each set of jitters, and subtracted before self-flattening the set. Although the 20 arcsec wide strips at the edges of the image have only one third of the exposure time, they have been retained with each source checked visually, and separate magnitude limits determined.
 
The combined $K$ image was convolved to the same seeing as the $I$ image, so that galaxy colours could be determined using a single fixed aperture, and then registered with the optical images. Photometric calibration of the $V$ and $I$ images onto the Landolt system was obtained using Landolt standard stars at varying airmasses, and a UKIRT faint standard was used for the $K$ image.
\begin{table}
\begin{center}
\begin{tabular}{|c|c|c|c|c|}\hline
Filter		& Exposure & Seeing   & \multicolumn{2}{|c|}{Completeness} \\
Band		& time	   & FWHM     & \multicolumn{2}{|c|}{limits (mag)} \\
(subfield)	& (sec)	   & (arcsec) & 90\%  & 50\% \\ \hline
$K$ (centre)	& 3780	   & 0.60     & 19.25 & 19.95 \\
$K$ (edge)	& 1260     & 0.60     & 18.82 & 19.53 \\
$I$		& 16200	   & 1.15     & 24.68 & 25.39 \\
$V$		& 8100     & 1.30     & 25.40 & 25.95 \\
\hline
\end{tabular}
\end{center}
\caption{Photometric properties of the images, including exposure times and 
completeness levels. The $V, I$ figures are for the area covered by the $K$ image, rather than for the whole BTC field.}
\label{photometry}
\end{table}

Object detection was carried out using {\sc SExtractor} \cite{bertin96} for objects with $\simeq{\rm FWHM}^{2}$ contiguous pixels over the $1\sigma$ detection threshold. Completeness levels for each filter (for the region covered by the $K$ image rather than the whole BTC field) are given in Table~\ref{photometry}. They were determined by dimming a bright galaxy to a specified magnitude and adding 100 copies to the image at random positions, and then processing through {\sc SExtractor} in the usual way. The 50\% and 90\% completeness levels are estimated as the magnitude for which 50 and 90 galaxies out of the 100 are recovered. The total magnitudes were taken to be the MAG\_BEST output from {\sc SExtractor}, and colours were determined using fixed apertures of diameter 2.5 arcsec.

\begin{figure*}[t] 
\psfig{figure=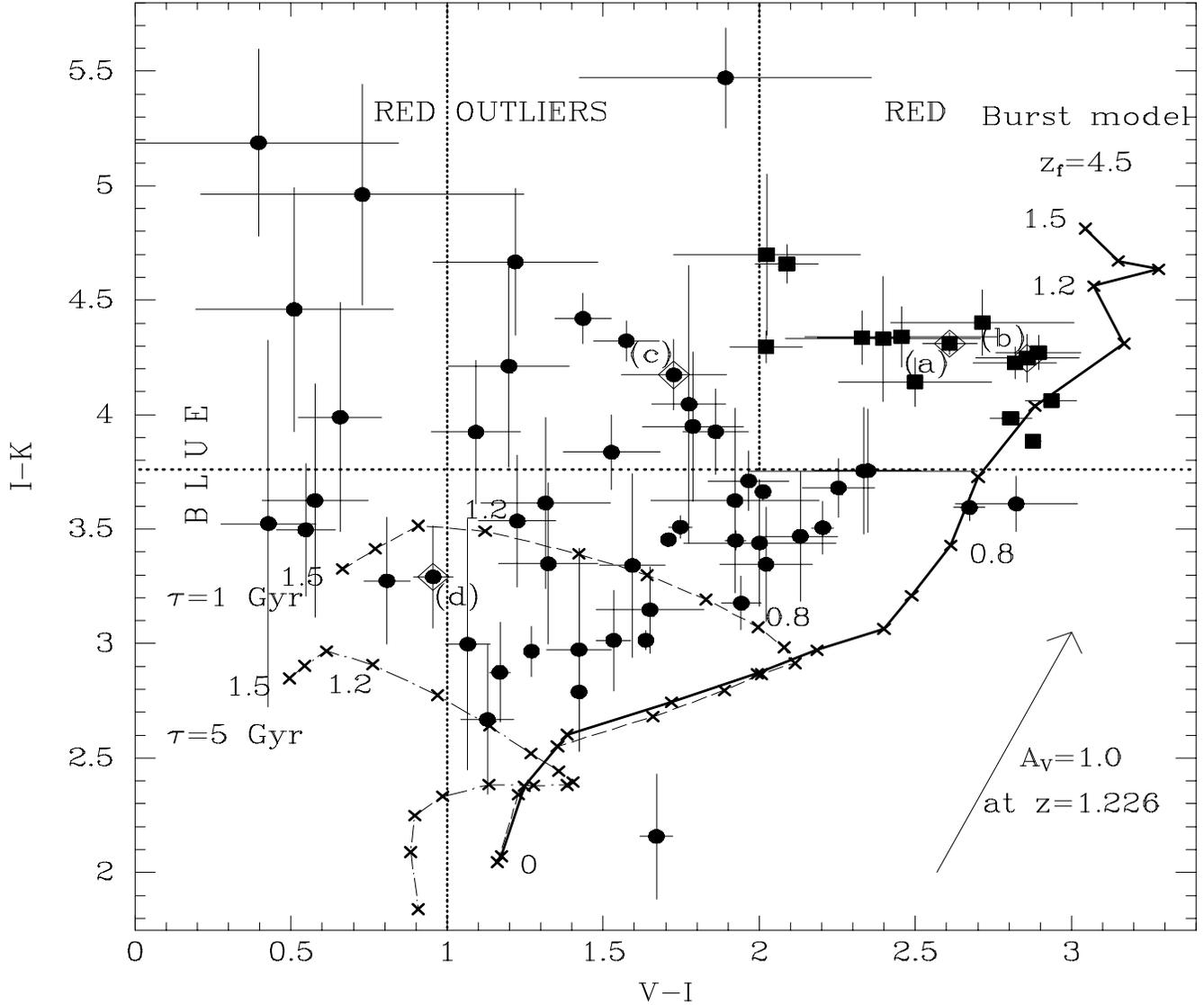,width=7.3in} 
\caption{$I-K$ against $V-I$ colour-colour diagram of all galaxies in the $K$ image. Those galaxies whose colours are well described by the burst model at $z\simeq 1.2$ are shown as filled squares. The four galaxies whose redshift probability distributions are shown in Fig.~\ref{zdist} are indicated by superimposed, labelled diamonds. For comparison, model tracks are shown for an instantaneous burst at $z_{formation}=4.5$ (solid line), and exponentially-decaying star-formation rate models with time scales of 1 Gyr (dashed line) and 5 Gyr (dot-dashed line). Each track shows the colour evolution from z=1.5 to z=0 with crosses at 0.1 redshift intervals. The effect of internal extinction at z=1.226 is shown as an arrow in the bottom-right corner. The dotted lines indicate the selection criteria used within this paper: the cluster candidates are selected to have $I-K>3.75$; with further separations into {\em red} ($V-I>2.00$) and {\em red outlier} ($V-I<2.00$) galaxies; {\em blue} galaxies are identified by $V-I<1.00$.}
\label{modelcols}
\end{figure*}

Star-galaxy separation of each source was performed using a combination of morphology (using the stellarity classifier in {\sc SExtractor}) and a comparison of $V-I$ and $I-K$ colours with model star and galaxy tracks. The stellarity classifier uses a neural net trained with a set of artificial stars and galaxies to produce a `probability' that a source is stellar. Stellarities of sources in both $V$ and $I$ images were obtained and the weighted mean used for classification. Most of the sources classified as stars morphologically, also had colours that lay near the model star tracks, but a number of sources had the colours of blue galaxies, and were reclassified as such.
\vspace{0.4in}

\section{Results}

\subsection{Galaxy Counts}
In total 100 visually-verified sources were detected in the $K$ image, of which 95 had counterparts in the $I$ image within 1 arcsec, and 79 had counterparts in both $V$ and $I$ images. 14 of the 79 sources were classed as stars, leaving 86 galaxies in the \mbox{$2.25\times2.25\,{\rm arcmin}^{2}$} field. Of the 5 sources detected in only the $K$ image, 2 are the close companion galaxies of the quasar which could not be separated from the quasar point-spread function by {\sc SExtractor} in the $V$ and $I$ images, and the 3 others must have $I-K>3.75$ to be undetected in $I$.

Table~\ref{GalaxyCounts} shows the comparison of galaxy counts seen in the $K$ image with those expected from a field region \cite{songaila94}. Excesses of galaxies are seen in all magnitude bins, even in the last bin where incompleteness should reduce any excess. If we limit ourselves to $K<19$ (an ${\rm L}_{K}^{*}$ galaxy at $z=1.226$ has $K\simeq19.05$) where we can be reasonably confident of both completeness and photometry, then a $3.5\sigma$ excess is observed (accounting for the effect on galaxy statistics of the two-point angular correlation function, $\omega(\theta)=1.13\theta('')^{-0.8}$ \cite{roche99}, for $K<19$ galaxies), with 40 galaxies observed whereas only 19 would be expected. By considering only those galaxies with $I-K>3.75$ then the excess is much more significant, with 23 galaxies observed instead of the expected 2. Even though the clustering amplitude of extremely-red galaxies is a factor ten greater than that of $K$-selected field galaxies \cite{daddi2000}, we still observe a $9\sigma$ excess. It is clear that the total excess is due entirely to these extremely-red galaxies, indicating a likely cluster at $z\gsim0.8$, and we consider all $I-K>3.75$ galaxies to be cluster members.

\begin{table}
\begin{center}
\begin{tabular}{|c||c|c||c|c|}\hline
& \multicolumn{2}{|c||}{Observed} & \multicolumn{2}{|c|}{Expected} \\ 
Magnitude    & Total & I$-$K$>$3.75   & Total & I$-$K$>$3.75 \\ \hline
$15<K\leq16$ & 1     & 1          & 0.533 & 0 \\
$16<K\leq17$ & 3     & 1          & 2.186 & 0 \\
$17<K\leq18$ & 13    & 5          & 4.204 & 0.050 \\
$18<K\leq19$ & 23    & 16         & 11.36 & 1.939 \\ 
$19<K\leq20$ & 39    & 23         & 21.81 & 1.896 \\ \hline
Total     & 79    & 46         & 41    & 4 \\ \hline
\end{tabular}
\end{center}
\caption{A comparison of total and extremely-red ($I-K>3.75$) galaxy counts in the $K$ image, binned by magnitude, and those expected for the same-sized region in the field (Songaila \etal 1994). The numbers include those observed only in the $K$ image, with the 2 companions to the quasar assumed to have \mbox{$I-K<3.75$}, and the other three sources to have $I-K>3.75$.}

\label{GalaxyCounts}
\end{table}

\subsection{Galaxy Colours and Photometric Redshift Estimates}

\begin{figure}
\psfig{figure=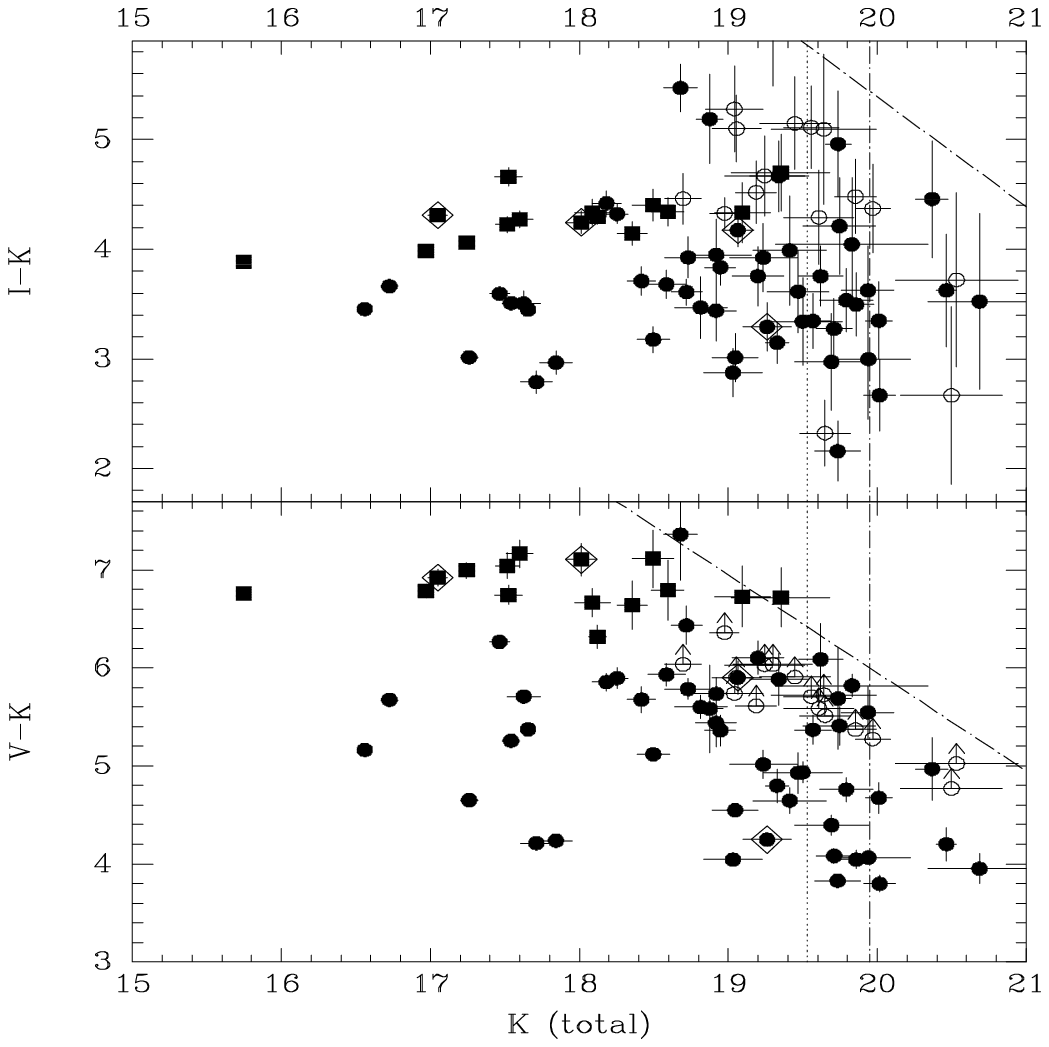,width=3.4in,height=4in}
\caption{Colour-magnitude diagrams of galaxies in the $K$ image. The solid symbols represent those galaxies detected in all three bands. The empty symbols represent those galaxies detected in $I$ and $K$ only. Galaxies whose colours are well described by the burst model at $z\simeq1.2$ have square symbols. The four galaxies whose redshift probability distributions are shown in Fig.~\ref{zdist} are indicated by diamonds. The dash-dotted lines indicate the 50\% completeness levels for each filter in the centre of the image, with the dotted line indicating the completeness level in $K$ for the edge of the image.}
\label{CMdiagram} 
\end{figure}

In Fig.~\ref{modelcols} we show the $I-K$ versus $V-I$ diagram for galaxies in the $K$ image. To estimate the photometric redshifts of each galaxy from its $VIK$ colours, and to produce the model colour tracks of Fig.~\ref{modelcols}, we have used the {\sc hyperz} code of Bolzonella, Miralles \& Pell\'{o} \shortcite{bolzonella2000}, which uses the Bruzual \& Charlot evolutionary code (GISSEL98, Bruzual \& Charlot 1993) to build synthetic template galaxies. It has stellar populations with 8 star-formation histories, roughly matching the observed properties of local galaxies from E to Im type: an instantaneous burst; six exponentially decaying SFRs with time-scales $\tau$ from 1 to 30 Gyr; and a constant star-forming rate. The models assume solar metallicity and a Miller \& Scalo IMF \shortcite{miller79}, with internal reddening considered through the Calzetti \etal \shortcite{calzetti2000} model with $A_{V}$ allowed to vary between 0 and 1 mag. The {\sc hyperz} software then produces a photometric redshift probability distribution through a chi-squared minimization process, allowing for all possible galaxy ages, star-formation histories and $A_{V}s$. This approach of determining a range of compatible redshifts, rather than a single best-fitting redshift for a galaxy, is more appropriate in this case given the limited colour information available. 

The model curves of Fig.~\ref{modelcols} correspond to stellar populations formed in an instantaneous burst at $z=4.5$ (solid line), and stellar populations with exponentially-decaying SFRs with time-scales ($\tau$) of 1 Gyr (dashed line) and 5 Gyr (dot-dashed line), and are thought to approximate the colour evolution of massive elliptical and disk-dominated galaxies. The galaxy ages for each model are calculated for an $h=0.5$, ${\rm q}_{0}=0.5$ universe. The $I-K$ colours of the model tracks increase monotonically with redshift to $z\simeq1.3$, and it is clear that $I-K\simeq4$ is a good indicator of galaxies at $z\gsim1$ with predominantly old stellar populations. The divergence of the disk and burst model tracks at high redshifts is a clear indication of the effect of recent star-formation on the $V-I$ colour in particular.
 
\begin{figure}
\psfig{figure=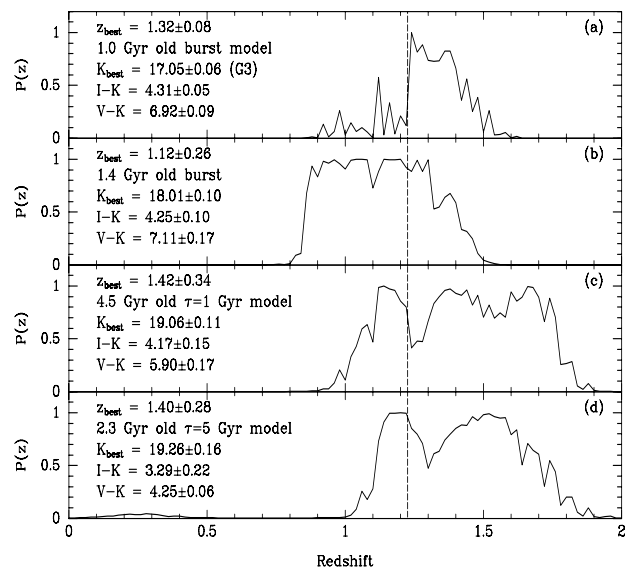,width=3.4in,height=3in}
\caption{Redshift probability distributions for 4 galaxies in the $K$ image based on their $VIK$ colours. The best fitting model is described by its redshift $z_{best}$ and star-formation history, which is either an instantaneous burst or an exponentially decaying star-formation rate with time-scale $\tau$. The vertical dashed line indicates the redshift of the quasar.
Distributions (a) and (b) correspond to two galaxies which help make up the observed red sequences in both $V-K/K$ and $I-K/K$ plots.
Galaxy (c) is a `red outlier' with a similarly red $I-K$ colour to (a) and (b), but a much bluer $V-I$ colour, indicating more recent star-formation. Galaxy (d) is one of the galaxies from the blue `band' and, due to its red $I-K$ colour, is best fit by a $z\gsim1$ disk galaxy. Note that the central dips in the probability distributions (c) and (d) are only artifacts due to the limited number of star-formation regimes available.}
\label{zdist}
\end{figure}

 In Figs~\ref{zdist}(a)--(d) we show the redshift probability distributions, P(z), for 4 galaxies which are compatible with being at the quasar redshift, incompatible with being at $z\lsim0.8$, and which represent the differing classes of galaxies discussed in the text below.

\begin{figure*}
\psfig{figure=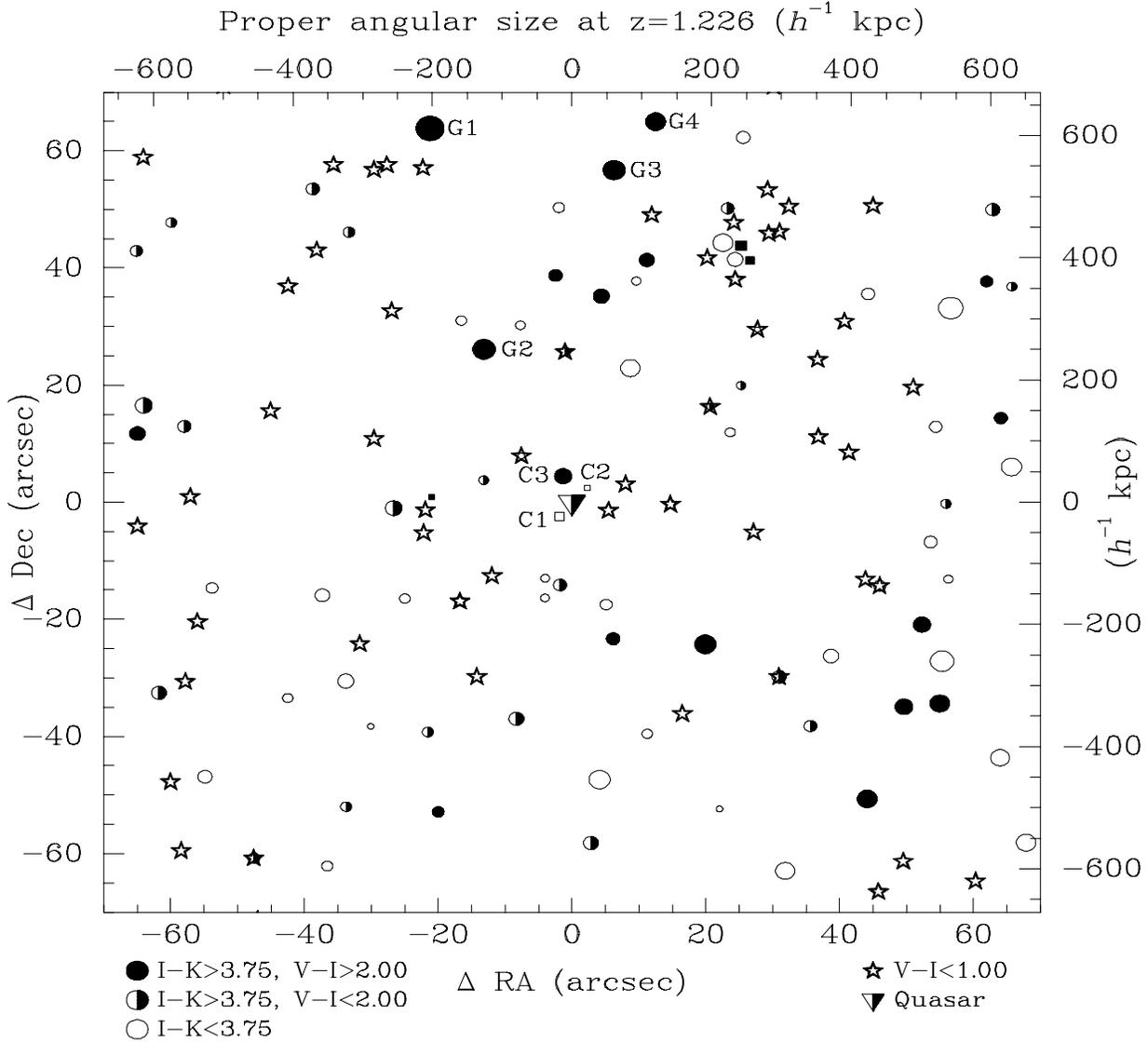,height=6.15in}
\caption{Spatial distribution of galaxies in the field of quasar 104420.8+055739. The different symbols indicate the galaxies' colour, with circles (squares) indicating those galaxies detected (not detected) in $I$. Solid symbols indicate the red ($I-K>3.75, V-I>2.00$) galaxies which could be early-type galaxies at the quasar redshift. Half-filled symbols indicate the red outlier ($I-K>3.75, V-I<2.00$) galaxies which are likely to be high redshift galaxies with some recent star-formation. The star symbols indicate the blue ($V-I<1.00$) galaxies which are probably undergoing significant star-formation. The size of the symbols (except stars) indicate the $K_{total}$ magnitude.}
\label{spatial}
\end{figure*}

\subsubsection{Red Sequence Galaxies}

There are 15--18 galaxies (hereafter labelled `red') whose extremely-red colours ($I-K>3.75, V-I>2.00$) are consistent with being passively-evolving galaxies at $z\simeq1.2$ (indicated by squares in Fig.~\ref{modelcols}). These red galaxies can be seen (as squares again) in the $V-K$ and $I-K$ against $K$ C-M diagrams of Fig.~\ref{CMdiagram} to form `fingers' at $V-K\simeq6.9$ and $I-K\simeq4.3$, comparable in form to the red sequences observed in lower redshift clusters, and in colour to other clusters at the same redshift. The $z\simeq1.2$ clusters of R99, S97, and D95 are all observed to have red sequences at $R-K\simeq5.9$, each with $\gsim4$ members spectroscopically confirmed as being at the cluster redshift. The red sequence galaxy colours can be compared directly for the R99 cluster where $I-K$ colour data exists, and the 4 spectroscopically confirmed red sequence members all have $4<I-K<4.4$, in good agreement with ours. The colours of these red sequence galaxies are well-fitted by the passively-evolving monolithic-collapse models of elliptical galaxies, and both their spectra and morphologies are similar to present-day ellipticals \cite{dickinson97}. Red sequence galaxies are usually the most luminous cluster members, the massive ellipticals, and this also appears to be the case here, with the 10 brightest galaxies with $I-K>3.75$ also having $V-K\approx6.9$. Given that these are both the reddest and most luminous galaxies, these should provide the tightest redshift estimates, and Fig.~\ref{zdist}a shows that the third brightest of these galaxies (G3) has \mbox{$1.2\lsim z\lsim 1.4$}. 

 The mean $I-K$ colour of the red galaxies is 4.25 with an intrinsic dispersion of 0.15 mag, and is best fit by a 2 Gyr old burst with a corresponding age dispersion of 400 Myr. The colour distribution seen in the red sequence is comparable to that seen in the cluster of T00 (0.22 mag in $R-K$), but is larger than that seen in the $z\simeq1.2$ 3C 324 cluster \cite{dickinson95} which has an rms scatter of 0.07 mag in $R-K$, suggesting that our cluster is less dynamically evolved than that of 3C 324.

\subsubsection{Red Outlier Galaxies}

Only half of the excess of $I-K>3.75$ galaxies is accounted for by the red sequence members, and we find comparable numbers (15--20) of $K\gsim19$ galaxies (hereafter labelled `red outliers') with both $I-K>3.75$ and $V-I<2.00$, which appear to fit neither the passively-evolving nor exponentially-decaying star-formation galaxy models (Fig.~\ref{modelcols}).

 Galaxies with similar colours and magnitudes have been observed in other $z\simeq1.2$ clusters (e.g. T00; Kajisawa \etal 1999; 2000).
 They have also been observed at $20\lsim K\lsim22$ in deep optical/NIR surveys (e.g. Moustakas \etal 1997), and appear common (several per square arcminute) at these fainter magnitudes. Much of the discussion (see e.g. Moustakas \etal 1997) of these objects has been limited to speculation due to lack of spectroscopic observations, but a widely held view is that they are probably high-redshift objects ($1\lsim z\lsim2$) which are undergoing significant star-formation and whose extremely-red $I-K$ colours are caused by a combination of dust and dominant old stellar populations. There is also some spectroscopic evidence that some $z\simeq1.2$ cluster ellipticals are undergoing star-formation, such as O\,{\sc ii} emission-lines seen in object \#4 of R99 and object \#237 of S97, both of which appear bluer in optical colours than the other red sequence galaxies in the clusters. Some extreme members of this population have been observed with $I-K>6$ (e.g. Hu \& Ridgway 1994), and one has since been spectroscopically confirmed as an ultraluminous infrared galaxy at z=1.44 with significant ongoing star-formation that is heavily obscured by dust (Graham \& Dey 1996; Dey \etal 1999). 
Given that both star-formation and dust are likely to have affected the colours of these `red outliers' significantly, it is not possible to constrain the galaxy redshifts beyond $z\gsim1$ (Fig.~\ref{zdist}c), but their prevalence in the vicinity of other high-redshift clusters, and the relative rarity of field $I-K>3.75$ galaxies, suggests that many are associated with the cluster.

\subsection{Spatial Distribution of Galaxies}

\begin{figure}
\psfig{figure=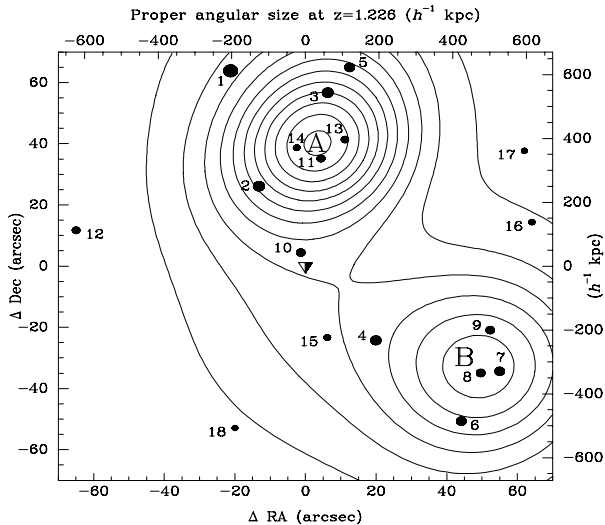,width=4.25in}
\caption{Estimated density distribution of the red galaxies in the field of quasar 104420.8+055739. The first contour corresponds to a density of 1 galaxy per square arcminute, and the separation of successive contours is also 1 galaxy per square arcminute. The quasar and red galaxies are marked as in Fig.~\ref{spatial}, and the labels correspond to the galaxy IDs of Table~\ref{probabilities}, which are numbered in order of increasing $K$ magnitude.}
\label{densitymap}
\end{figure}

The spatial distribution of galaxies detected in the $K$ image is shown in Fig.~\ref{spatial}, with solid symbols used to differentiate the red galaxies from the remainder.

\subsubsection{Red Sequence Galaxies}

\begin{figure}
\psfig{figure=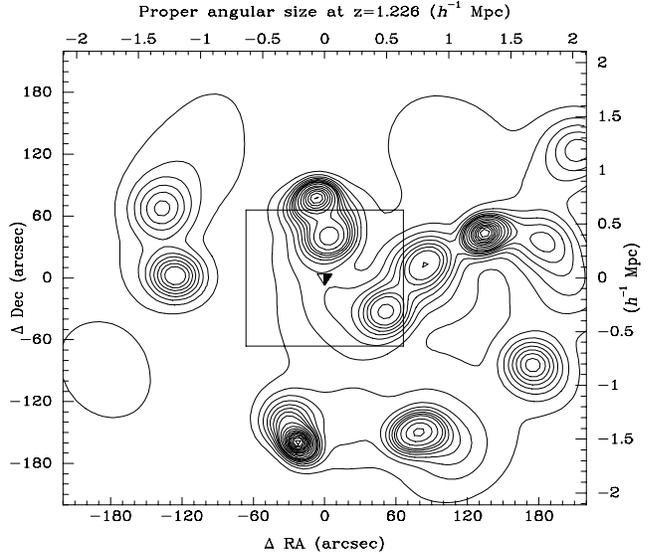,width=4.5in}
\caption{Contour plot of the estimated density distribution of optically-red  ($V-I\gsim2.25,I<23$) galaxies (including all C-M relation members) in a $7\times7\,{\rm arcmin}^{2}$ (corresponding to $4.06\times4.06\,h^{-2}\,{\rm Mpc}^{2}$ at z=1.226) field centred on the quasar. The first contour corresponds to a density of 1 galaxy per square arcminute, and the separation of successive contours is also 1 galaxy per square arcminute. The quasar is marked as previously, and the central box corresponds to the field of the $K$ image.}
\label{lss}
\end{figure}

 The red galaxies appear to be distributed across the $K$ image (\mbox{$1.29h^{-1}{\rm Mpc}$} at z=1.226) with no concentration towards the quasar. However, the ten most luminous red galaxies (which we indicated earlier are probably massive ellipticals at the quasar redshift) are concentrated in two compact groups, one towards the top-centre of the image (along with a number of fainter members), and the other in the south-eastern corner. This suggests that the galaxy excesses and red sequences could be due to two clusters at similar redshifts.
 Assuming that the brightest of the red galaxies, G1, is a quiescent galaxy at the quasar redshift, then it has \mbox{${\rm L}\approx14{\rm L}_{K}^{*}$}, and is more luminous in $K$, by almost a magnitude, than any of the galaxies from other $z\sim1.2$ clusters (D95; S97; R99). It is also a radio emitter, being the only source in the $K$ image detected by the VLA FIRST 20cm survey, having an integrated flux of $2.63\pm0.15$mJy, which suggests that it has an active nucleus. It is common for the brightest cluster galaxy (which is what we assume this is) to also be a radio source, although it is not as spectacular an example as 3C 324 which is at a similar redshift, but is 1000 times more luminous in the radio. If this galaxy is comparable to 3C 324 then it may also display narrow-line emission, in particular O{\sc ii} and Mg{\sc ii}, and this may explain why it appears bluer than the other red sequence galaxies, as the emission-lines boost the optical flux. The brightest cluster galaxy is usually located near the cluster centre, and it appears to be the case here too if Figs.~\ref{spatial} and~\ref{lss} are compared. 

To examine the significance of any substructure for the red galaxies the non-parametric cluster analysis method described by Pisani (1993,1996) is applied.
 This is based upon estimating the underlying probability density, $F(\xvec)$, of a data sample \mbox{$D_{N}=\{x_{1},x_{2},...,x_{N}\}$} using an iterative and adaptive kernel method. 
 Each data point, $x_{i}$, is initially smoothed by a Gaussian kernel of width $\sigma$ to produce a pilot fixed-kernel estimate of the local density, $f_{p}(x_{i})$, at each point. The optimal value for the smoothing parameter $\sigma$ is determined by minimizing the integrated square error, $ISE(f_{p})$, of the estimation of $F(\xvec)$. It is possible to show (see Silverman 1986) that minimizing the cross-validation, $M(f_{p})$, a function related to $ISE(f_{p})$, and that can be written as a sum of functions dependent only on the $x_{i}$, is equivalent to minimizing $ISE(f_{p})$.  The pilot local density estimate is then used to adapt the amount of smoothing applied to each point through \mbox{$\sigma_{i}=\sigma [f_{p}(x_{i})/\bar{f}_{p}]^{-1/2}$}, where $\bar{f}_{p}$ is the geometric mean of the $f_{p}(x_{i})$. The adaptive kernel estimate of the local density $f_{a}(\xvec)$ is then

\begin{equation}
f_{a}(\xvec)=\frac{1}{N}\sum_{i=1}^{N} \frac{1}{2\pi\sigma_{i}^{2}}\exp \left[-\frac{1}{2}\frac{|{\bf x}-x_{i}|^{2}}{\sigma_{i}^{2}}\right].
\end{equation}
 The adaptive kernels produce both a higher resolution in the clustered regions where it is needed, and increased smoothing in the low-density regions (see Silverman 1986; Pisani 1993;1996 for detailed discussion), and the non-parametric nature of the method means that the amount of smoothing is dependent solely upon the data points rather than any prior estimate of what the cluster width should be. The probability density estimate of red galaxies is shown in Fig.~\ref{densitymap} with the two density peaks, marked A and B, which are assumed to mark the two cluster centres.

Having produced the probability density estimate, each galaxy is assigned membership to one of the clusters by following a path from the original position $x_{i}$ along the maximum gradient of $f_{a}(\xvec)$ until it reaches the local maximum A or B. The contribution to the local density from each cluster, $f_{A}(\xvec),f_{B}(\xvec)$, is then taken to be the sum of the kernels for galaxies assigned to that cluster.

\begin{table}
\begin{center}
\begin{tabular}{|c|c|c|c|c|c|c|}\hline
ID	& $K_{tot}$	& A/B & $P(i\in 0)$	& $P(i\in A)$	& $P(i\in B)$	\\ \hline
1  & 15.75 & A & 0.0827 & 0.9075 & 0.0098 \\
2  & 16.97 & A & 0.0324 & 0.9259 & 0.0417 \\
3  & 17.05 & A & 0.0259 & 0.9626 & 0.0115 \\
4  & 17.24 & B & 0.0562 & 0.1172 & 0.8266 \\
5  & 17.52 & A & 0.0391 & 0.9468 & 0.0142 \\
6  & 17.52 & B & 0.0438 & 0.0128 & 0.9434 \\ 
7  & 17.60 & B & 0.0319 & 0.0159 & 0.9522 \\
8  & 18.01 & B & 0.0306 & 0.0164 & 0.9529 \\
9  & 18.09 & B & 0.0357 & 0.0298 & 0.9345 \\
10 & 18.12 & A & 0.0583 & 0.6754 & 0.2663 \\
11 & 18.35 & A & 0.0188 & 0.9582 & 0.0230 \\
12 & 18.50 & A & 0.6440 & 0.2716 & 0.0844 \\
13 & 18.59 & A & 0.0200 & 0.9595 & 0.0205 \\
14 & 18.95 & A & 0.0191 & 0.9641 & 0.0167 \\
15 & 18.97 & B & 0.0732 & 0.2129 & 0.7139 \\
16 & 19.09 & B & 0.1860 & 0.2110 & 0.6030 \\
17 & 19.24 & A & 0.2852 & 0.3465 & 0.3683 \\
18 & 19.35 & B & 0.2512 & 0.1011 & 0.6477 \\
\hline
\end{tabular}
\end{center}
\caption{Results from the cluster membership analysis of the red galaxies. Column A/B indicates whether a galaxy was assigned to cluster A or B, and the last three columns show the probabilities that a galaxy is isolated or a member of cluster A or B.}
\label{probabilities}
\end{table}

 The expected background contamination from red field galaxies, $f_{0}(\xvec)$, is unknown, and so we have estimated it to be half that of the $I-K>3.75$ field galaxies or 0.5 galaxies per square arcminute, as about half our $I-K>3.75$ galaxies are also classified as red (\mbox{$I-K>3.75,V-I>2.00$}).
 The probability that each galaxy is isolated, and hence due to the background component, is just the fractional contribution of the background density to the local density \mbox{$P(i\in 0)=f_{0}/f_{a}(x_{i})$}. The probability that galaxy $i$ is a member of cluster $A$ is then the fractional contribution of cluster A to the local density, after considering the background contribution \mbox{$P(i\in A)=(1-P(i\in 0))f_{A}(x_{i})/f_{a}(x_{i})$},
 and for each of the red galaxies these probabilities are shown in Table~\ref{probabilities}. Looking at this table and Fig.~\ref{densitymap} it appears that the two clusters are well defined, with 11 out of the brightest 14 galaxies having probabilities  greater than 0.90 of having been assigned correctly to cluster A or B. Only galaxies \#4 and \#10 have significant doubt over cluster assignment, and \#12 appears to be isolated, whilst the four faintest red galaxies show no clear membership of either cluster, and may only be part of the combined structure. 

As a second estimate of the significance of the substructure, we examine the null hypothesis that the galaxies are all members of a single cluster. One method for examining the null hypothesis is the likelihood ratio test statistic (LRTS) (Ashman, Bird \& Zeff 1994; Kreissler \& Beers 1997) which evaluates the improvement in fitting the data of a two-component model over a single elliptical Gaussian probability density function. The best-fit single elliptical Gaussian $f_{(1)}(\xvec)$ to the data is found using the mean, $\bar{\xvec}$, and covariance matrix, {\bf A}, of the data, and this unimodal probability density function is used to create 10000 bootstrap catalogues of 18 galaxies each. The adaptive kernel estimator is applied to each bootstrap catalogue in the same way as the original data, and for those catalogues where bimodality is observed, the cluster membership probabilities $P(i\in\mu)$, for $i=1,...,N;\mu=1,2$ are calculated. The probable number of galaxies in cluster $\mu$ is then given by \mbox{$n_{\mu}=\sum_{i=1}^{N}P(i\in\mu)$}. The best-fit double Gaussians, $f_{(g;\mu)}(\xvec)$, can then be found using the means, $\bar{\xvec}_{\mu}$, and covariance matrices, ${\bf A}_{\mu}$, of the dataset weighted by $P(i\in\mu)$ for $\mu=1,2$, before normalizing to $\int_{\Re}f_{(g;\mu)}(\xvec)d\xvec=n_{\mu}$. The measure of the fit by the g-component model is evaluated through the likelihood value
\begin{equation}
L_{C}(g)=\prod_{i=1}^{N} \left( \sum_{\mu=1}^{g} f_{(g;\mu)}(x_{i})\right)^{1-P(i\in0)}.
\end{equation}
 The evaluation of the improvement in going from a single to a double Gaussian fit is then given by the LRTS $\lambda=-2\ln\{L_{C}(2)/L_{C}(1)\}$. The significance of the substructure observed is estimated by the probability that a null hypothesis bootstrap catalogue produces a value of $\lambda$ greater than the observed value. For the observed distribution, using 10000 bootstrap catalogues, we find that the substructure is marginally inconsistent with the null hypothesis at the 8.88\% significance level. The low level of significance is due to the small number of galaxies involved, as many of the more significant substructures among the bootstrap catalogues were due to just 3 or 4 points that were within a few arcsec of one another.

\subsubsection{Clustering across Large-Scales}

As both groups are located near the edges of the $K$ image, and so may suffer from truncation, the full extent of clustering associated with the quasar is estimated by considering a $7\times7\,{\rm arcmin}^{2}$ field centred on the quasar from the BTC $V$ and $I$ images. Fig.~\ref{modelcols} indicates that any $z\gsim1$ quiescent galaxies should be amongst the reddest galaxies in $V-I$, and so by selecting those galaxies with $V-I>2.25$ the density contrast due to any clustering at $z\simeq1.2$ should be maximised. Note that this selection criterion is slightly redder than the $V-I>2.00$ selection used previously, in order to reduce the contamination of intermediate-redshift quiescent galaxies. In the region covered by the $K$ image, only the red galaxies are selected, including the 3 with $2.00<V-I<2.25$. The estimated density distribution of $V-I>2.25,I<23$ galaxies is shown in Fig.~\ref{lss}, and it is clear that the clustering extends well beyond the $K$ image. Cluster A now appears to be centred on the northern edge of the $K$ image, and cluster B appears part of an elongated structure which extends 2--3 arcmin to the north-east. Two further groups, each of 6--10 optically-red galaxies, are apparent 3 arcmin south of the quasar and 1.5 arcmin apart. The significance of the substructure formed by clusters A and B appears to be much higher now, as the density peaks are moved further apart. A re-examination of the null hypothesis, after including the $V-I>2.25$ galaxies outside the K image, finds the substructure to be inconsistent with the null hypothesis at the 1.86\% significance level. The increase in significance suggests that the clusters were truncated by the boundaries of the $K$ image, but the increase is also due partly to the larger galaxy sample.

\subsubsection{Companion Galaxies to the Quasar Host}

There are two galaxies (labelled C1 and C2 in Fig.~\ref{spatial}) which are only 3 arcsec $(12.9\kpc)$ from the quasar, and could well be companion galaxies to the quasar galaxy host. Compact companions are found for a significant fraction of quasars (e.g. Bahcall \etal 1997) and spectroscopic observations confirm that many have stellar populations and redshifts within 500 km ${\rm s}^{-1}$ of the quasar (e.g. Stockton 1982; Canalizo \& Stockton 1997). It has been suggested (e.g. Stockton 1982; Bekki 1999) that these companion objects are tidally-stripped cores from galaxies that have recently interacted with the quasar host galaxies, and that this interaction provides an efficient fuelling mechanism for quasar activities. Spectroscopic analysis of the companion to quasar PG 1700+518 (Canalizo \& Stockton 1997) finds evidence for both a starburst event that occurred roughly 100 Myr ago (and so could be coincident with the quasar activation), and a relatively old stellar population likely to be from the merger progenitor disk. The galaxies C1 and C2 are bluer than the red sequence galaxies (cf. C3), and so, if associated with the quasar, they have undergone a recent episode of star-formation, presumably caused by the merger process.

\subsubsection{Blue Galaxies}

To compare this quasar field with the results of HCJ we have also examined the distribution of blue ($V-I<1$) galaxies for $I<25$ (indicated by star symbols in Fig.~\ref{spatial}). We find a concentration of blue galaxies within 30 arcsec $(290\kpc)$ of the quasar, which appears to be extended towards the north-east, forming a `band' that bisects the two groups of red galaxies. The band presumably corresponds to a region of enhanced star-formation. Few, however, are found near the centres of either group of red galaxies. In comparison with adjacent fields (over a \mbox{$7'\times7'$} region) in the optical images we do not find an excess of blue galaxies. 

\begin{figure}
\psfig{figure=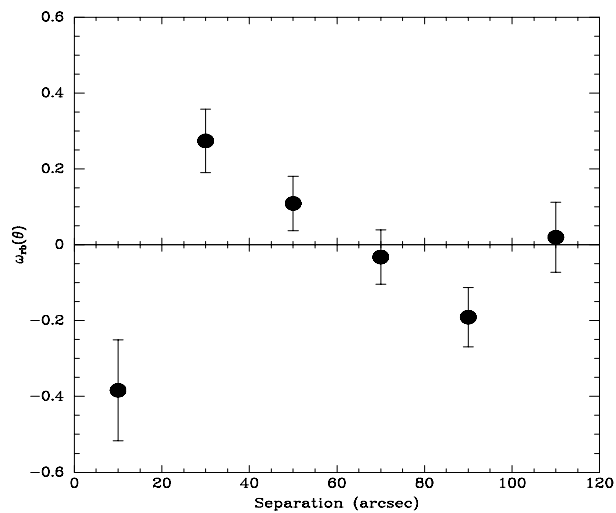,width=4.25in}
\caption{The angular cross-correlation, $w_{rb}(\theta)$, between red ($V-I>2.00,I-K>3.75$) and blue ($V-I<1.00$) galaxies in the field covered by the $K$ image. The errors are assumed to be Poissonian with variance $N_{rR}$.}
\label{twopoint}
\end{figure}

To examine the relative spatial distribution of red and blue galaxies, the angular cross-correlation function, $\omega_{rb}(\theta)=N_{rb}(\theta)/N_{rR}-1$, is determined (see Fig.~\ref{twopoint}), where $N_{rb}(\theta)$ is the number of red-blue galaxy pairs with separation $\theta$, and $N_{rR}(\theta)$ is the expected number of pairs with one member from the red catalogue and the other from one of 10000 randomly distributed catalogues.

At small separations ($\theta<20''$) the red and blue galaxies are strongly anti-correlated at the $3\sigma$ level, confirming the apparent avoidance of the red groups by the blue galaxies in Fig.~\ref{spatial}.
 In contrast, for $20''<\theta<60''$, there is a correlation at the $3\sigma$ level between red and blue galaxies, which is due to the `band' of blue galaxies that bisects the two groups of red galaxies.

 The blue galaxies do not contribute to the excess of galaxies observed at $K<19$, as only 9 of 51 are detected in $K$, the brightest having $K=18.87$. These 9 all have the red $I-K$ colours indicative of $z\gsim1$ galaxies (e.g. Fig.~\ref{zdist}d), of which 6 help make up the concentration $\lsim$30 arcsec from the quasar, which suggests that this at least is real. However, given that the $V-I$ colour is affected much more by recent star-formation than redshift, we cannot say anything about the likely redshift of the remaining blue galaxies.


\section{Discussion and Conclusions}

We have found a $3.5\sigma$ excess of $K<19$ galaxies in the \mbox{$2.25\times2.25\,{\rm arcmin}^{2}$} field around the z=1.226 radio-quiet quasar 104420.8+055739 from the Clowes-Campusano LQG. The excess is due entirely to a factor $\sim$11 overdensity of red ($I-K>3.75$) galaxies, which must have $z\gsim0.8$ to explain their colour. In particular, we find cluster red sequences of 15--18 galaxies, in the $I-K$ and $V-K$ against $K$ C-M diagrams at $I-K\simeq4.3, V-K\simeq6.9$, comparable in both colour and magnitudes to red sequences observed for other $z\simeq1.2$ clusters. These red sequences suggest a population of massive ellipticals at the quasar redshift. In the area of the $K$ image we find tentative evidence of substructure amongst these galaxies, with two apparent groups 40 arcsec to the north and 60 arcsec to the south-east of the quasar. An examination of the optical images over a \mbox{$7\times7\,{\rm arcmin}^{2}$} area indicates that this substructure is significant at the 2\% level, and that the clustering extends well beyond the $K$ image, forming a large-scale structure \mbox{2--3$\mpc$} across. The overall structure is suggestive of being the early stages of formation of a cluster from the progressive coalescence of subclusters. 

Only half of the excess of $I-K>3.75$ galaxies is accounted for by the red sequence members, and we find comparable numbers (15--20) of `red outlier' galaxies  with both $I-K>3.75$ and $V-I<2.00$, which appear to fit neither the passively-evolving nor exponentially-decaying SFR galaxy models (Fig.~\ref{modelcols}).  Although, all we can say is that they are likely to be dusty star-forming galaxies at $1\lsim z\lsim2$, given that such galaxies are found around other $z\simeq1.2$ clusters (e.g. T00; Kajisawa \etal 1999; 2000), and given their comparative rarity in field regions, it seems reasonable to assume that they are associated with the cluster. This would suggest that the Butcher-Oemler effect observed in intermediate-redshift clusters increases in strength to higher redshifts, with $\gsim50\%$ of likely cluster members exhibiting the blue colours of recent star-formation.

We also find a concentration of blue ($V-I<1$) galaxies within 30 arcsec $(130\kpc)$ of the quasar, with many having the red $I-K$ colours of $z\gsim1$ galaxies, which suggests that the quasar lies in a region of enhanced star-formation, in agreement with the results of HCJ. This concentration appears extended in such a way as to separate the two groups of red galaxies, and it is also notable how the blue galaxies appear to avoid the centres of red galaxy clustering. Some foreground contamination is likely, but given that such concentrations appear common around LQG quasars at this redshift, and the highly significant spatial interrelation between the blue and red galaxies, then we can be reasonably confident that this concentration is real, and is associated with the quasar.
 However spectroscopic observations of these and the other cluster candidates will be required to confirm cluster membership, and to provide more quantitative information about their star-formation histories, such as their approximate ages and current star-formation rates.

\subsubsection{A Possible Cluster Merging Event}

The relative distribution of red and blue galaxies can be explained if what we are witnessing is the early-stages of merger of the two clusters of red galaxies, which has triggered both the band of enhanced star-formation and the quasar itself. A comparable distribution has been observed for the Coma cluster \cite{caldwell93} with a band of post-starburst or `E+A' galaxies located between the Coma cluster centre and a secondary X-ray peak. Dynamical studies \cite{burns94} indicated that these galaxies had passed through the centre of the Coma cluster about 2 Gyr ago, coincident with the epoch of starbursting predicted from the spectra of the post-starburst galaxies. Unusually high blue galaxy fractions have been observed for a number of low-redshift clusters with bimodal X-ray surface brightness profiles \cite{metevier2000}, implying that cluster mergers can induce starbursts simultaneously in a large fraction of cluster galaxies, and they could be a major contributor to the Butcher-Oemler effect. Several mechanisms have been suggested that could cause firstly the triggering and then the termination of a secondary burst of star-formation in a galaxy, as a subcluster passes through a cluster. These include ram pressure from the ICM \cite{evrard91}, shocks due to collisions between the two ICMs \cite{roettiger96}, and the effect of close galaxy encounters \cite{moore96}. As the `band' of blue galaxies is likely to be undergoing or has recently undergone star-formation, and because the two groups are relatively close together, we suggest that this system is being observed at an earlier epoch of the cluster merger process than Caldwell \etal (1993), either just before or just after core passage, and that the star-formation has been triggered by the interaction of galaxies with the shock fronts produced by the collision of the two ICMs (see Roettiger \etal 1996).

Cluster merging events are predicted to be relatively common at high redshifts ($z\gsim1$) in hierarchical clustering models (e.g.
 Press \& Schechter 1974;
 Bahcall, Fan \& Cen 1997;
 Percival \& Miller 1999).
 Examples of possible merging clusters at high redshifts are the CL0023+00423 groups at z=0.8274 and 0.8452 which, according to a dynamical study, have a 20\% chance of merging \cite{lubin2000}, and the R99 and S97 clusters which are separated by only $2.5\mpc$.
 
\subsubsection{Comparison with Other Work}

In a study of 7 radio-loud quasars at $1.0<z<1.6$, S\'{a}nchez \& Gonz\'{a}lez-Serrano (1999) find excesses of faint ($B>22.5$ and $R>22.0$) galaxies on scales of $r<170$ arcsec and $r<35$ arcsec around the quasars, whose numbers, magnitudes and angular extensions are compatible with being clusters of galaxies at the quasar redshifts. In particular however, they find that the quasars are in general not located at the peaks of the density distribution, but are some 40--100 arcsec from them, located on the cluster peripheries. This result, and our own, is understandable in the framework of quasar activity being triggered by the infall of gas onto a seed black hole. Firstly, galaxies in the centres of clusters have previously lost most or all of their gas, by having had it stripped off by ram-pressure from the intra-cluster medium or by tidal forces from close encounters with other galaxies, or by consuming the gas in a starburst during its first infall into the cluster. Secondly, the encounter velocities for galaxies in the cluster cores greatly exceed the internal velocity distribution of the galaxies, making galaxy mergers much less effective at triggering nuclear activity \cite{aarseth80}.

In a comparable study of the galaxy environment of the radio-loud quasar 1335.8+2834 from the $z\sim1.1$ Crampton \etal (1989) LQG, Tanaka \etal (2000a) obtain results which have several similarities to our own. This quasar had been part of the HCJ study, and was known to lie in a band of blue and emission-line galaxies. Using deep $R,I$ and $K$ observations, a number of extremely-red objects with the colours of passively-evolving galaxies at the quasar redshift were found, forming a cluster which lies to one side of both the quasar and the band of blue and emission-line galaxies. They also find a similar population of `red outliers' and estimate the blue galaxy fraction as 60--80\%. They also find an indication that this cluster is part of a larger structure with groupings of optically-red (\mbox{$R-I>1.3$}) galaxies, similar to the clustering near the quasar, found across the \mbox{$8\times8\,{\rm arcmin}^{2}$} (\mbox{$4.7\times4.7\,h^{-2}\,{\rm Mpc}^{2}$} at z=1.086) $R,I$ optical images. 

In a wide-field ($48\times9\,{\rm arcmin}^{2}$) optical imaging survey of the 1338+27 field containing 5 quasars from the Crampton \etal (1989) LQG, Tanaka \etal (2000b) detect significant clustering of faint red galaxies with $I>21$ and \mbox{$R-I>1.2$}. These galaxies are concentrated in 4--5 clusters forming a linear structure of extent $\sim10\mpc$ that is traced by the group of quasars, although only the one radio-loud quasar of T00a appears to be directly associated with any of the rich clusters. The immediate environments of the other four LQG quasars, all radio-quiet, appear relatively poor in terms of red galaxies, although three have excesses of blue and emission-line galaxies (HCJ95) indicating that they are located within regions of enhanced star-formation. This is the clearest evidence yet that LQGs trace large-scale structure, even if the majority of the member quasars are only directly associated with regions of enhanced star-formation, rather than rich clusters.

\subsubsection{Interpretation - Mechanisms for Quasar Formation}

As several of the HCJ quasars were found in regions of enhanced star-formation, and as both the quasar of T00a and this paper are located in `bands' of enhanced star-formation, in between, or on the peripheries of, clusters, we propose a causal link between the quasar and star-formation whereby both are triggered by the same mechanism: the interaction between the galaxy and the intra-cluster medium. 
If a galaxy can be disrupted sufficiently by its passage through the ICM to cause it to undergo starbursting, then if it also contains a supermassive black hole, enough gas may be channelled onto the nucleus to trigger a phase of quasar activity. Such a mechanism explains the finding of quasars in regions of enhanced star-formation more naturally than the galaxy merger model, as it allows many galaxies to be affected simultaneously, although there is good evidence that a large fraction of quasars have been triggered by galaxy mergers. If there is a connection between the quasar and star-formation activation, then it is likely that these quasars are found preferentially in clusters with high blue galaxy fractions, and that these are not representative of $z\simeq1.2$ clusters as a whole.

This work and previous studies show that searching for sources with the $I-K>3.75$ colours characteristic of quiescent galaxies at $z\gsim1$ is an efficient means of locating $z\gsim1$ clusters. By adding a second optical band it is then possible to obtain qualitative information on the star-formation history of these galaxies. Using this information along with the relative spatial distribution of quiescent and star-forming galaxies, a more complex picture arises in which the evolution of galaxies, quasars and clusters are all interrelated.

\section{Acknowledgements}

The optical data were obtained with the CTIO 4-m Blanco telescope, and the data reduced with IRAF and the UK STARLINK facilities. Advice regarding the removal of distortions from the BTC images was given by Gary Bernstein and Tony Tyson. Roser Pell\'{o} provided the photometric redshift estimation software {\sc hyperz}, and gave advice regarding its use and models. CPH acknowledges a PPARC studentship, and the Universidad de Chile and CTIO for their hospitality during his stay in Chile. LEC was partially supported by FONDECYT grant 1970735. We thank the referee for helpful comments regarding the statistics.

IRAF is distributed by the National Optical Astronomy Observatories which is operated by the Association of Universities for Research in Astronomy, Inc. (AURA) under cooperative agreement with the National Science Foundation.


\end{document}